\documentstyle[amsmath,multicol,epsfig,prb,aps,color,graphicx]{revtex}

\begin{document}
\newcommand{\Eq}[1]{Eq.~(\ref{#1})}
\newcommand{\fg}[1]{Fig.~\ref{#1}}
\newcommand{\cph}[1]{Chem. Phys. {\bf #1}}
\newcommand{\prevb}[1]{Phys. Rev. B{\bf {#1}}}
\newcommand{\prevl}[1]{Phys. Rev. Lett. {\bf {#1}}}
\newcommand{\sm}[1]{Synth. Metals {\bf {#1}}}
\newcommand{\jchemp}[1]{J. Chem. Phys. {\bf {#1}}}
\newcommand{\jphysc}[1]{J. Phys. Chem. {\bf {#1}}}
\newcommand{\jpcb}[1]{{J. Phys. Chem. B {\bf #1}}}
\newcommand{\pcp}{$\pi$-conjugated polymer}
\newcommand{\red}{\color{red}}
\newcommand{\blue}{\color{blue}}
\newcommand{\w}{\mbox{$\omega$}}
\newcommand{\D}{\mbox{$\Delta$}}
\newcommand{\G}{\mbox{$\Gamma$}}
\newcommand{\smq}{\mbox{$\simeq$}}
\newcommand{\ee}{E. Ehrenfreund}
\newcommand{\zvv}{Z.V. Vardeny}
\newcommand{\bi}{\bibitem}
\newcommand{\beq}{\begin{equation}}
\newcommand{\eeq}{\end{equation}}
\newcommand{\technion}{Technion--Israel Institute of Technology,
Haifa 32000, Israel}
\newcommand{\ssi}{Solid State Institute}
\newcommand{\phy}{Department of Physics}
\newcommand{\hs}{\hspace{0.4cm}} 
\newcommand{\npi}{\hspace{-0.4cm}} 
\newcommand{\vs}{\vspace{0.4cm}} 

\newcounter{fg}
 \refstepcounter{fg} \label{PL} \refstepcounter{fg} \label{Raman}
\refstepcounter{fg} \label{model}

\title{
 Apparent vibrational side-bands in $\pi$-conjugated
systems: the case of distyrylbenzene}

\author{
C.C. Wu$^1$, \ee$^{1,2}$, J.J. Gutierrez$^3$, J.P. Ferraris$^3$,
\zvv$^{1}$}
\address{$^1$\phy, University of Utah, Salt Lake City,
UT 84112\\
$^2$\phy\ and \ssi, \technion\\
$^3$ Nano Tech Institute, University of Texas at Dallas,
Richardson, TX 75083
 }

\maketitle

\begin{abstract}

The photoluminescence (PL) spectra of dilute solution and single
crystals of distyrylbenzene show unique temperature dependent
vibronic structures. The characteristic single frequency
progression at high temperatures is modulated by a low frequency
progression series at low temperatures. None of the series side
band modes corresponds to any of the distyrylbenzene Raman
frequencies. We explain these PL properties using a time dependent
model with temperature dependent damping, in which the many-mode
system is effectively transformed to two- and then to a single
"apparent" mode as damping increases.

\end{abstract}

\setlength{\columnsep}{6mm}
\begin{multicols}{2}

\npi The excited state properties of distyrylbenzene (DSB), which
is the three phenyl group oligomer analog of p-phenylene vinylene
(\fg{PL}, inset), have been the subject of recent experimental
\cite{obl98,mcsb02} and theoretical \cite{span01} spectroscopic
studies, because of potential opto-electronic applications
\cite{Hadz}. DSB photoluminescence (PL) spectroscopy has been the
subject of numerous research studies, since it is strongly
dependent upon the packing order
\cite{span01,oegt96,gmlo02,wdvf03}. In DSB films the molecules
form H aggregates thus substantially weakening the emission
quantum yield, relative to the separate oligomers in dilute
solutions \cite{oegt96}. This is especially true in DSB single
crystals: due to the herringbone symmetry, the fundamental optical
transition (the so called "0-0 transition") is either totally
absent or significantly reduced \cite{span01,oegt96}. Typical PL
spectra of DSB chromophores contain vibronic progression series,
of which the relative intensity and frequencies depend on packing
and temperature. One key feature of the vibronic progression is
that the side-band replicae have apparent frequencies that do not
match {\bf any} of the Raman active modes. This is true also for
many other \pcp s and oligomers: the rich Raman spectrum of the
multi-vibrational modes systems is reduced to only one, or two
apparent vibronic progression in their PL spectrum.

\npi In this work we account for the apparent modes that appear in
the PL spectra of DSB dilute solutions and single crystals; the
same model may be used to explain the PL spectra of other
$\pi$-conjugated systems. We use the pre-resonance Raman spectrum
in order to quantify the relative configuration displacement (or,
equivalently, the Huang-Rhys (HR) factor) for each of the
vibrational modes. Consequently, we utilize the relative HR
factors in a damped time-dependent model to calculate the DSB PL
spectrum. We found that above a certain finite damping, the PL
vibronic progression contains only a single apparent vibronic
frequency. For lower damping, however, the single frequency
progression is modulated by a low frequency progression, in
excellent agreement with the PL spectra measured at low
temperatures.

\npi In \fg{PL}a we show the PL spectra of DSB in dilute frozen
solution of tetra-decane \cite{gmlo02} at low and high
temperatures.
 The PL spectrum at 200 K has the typical vibronic
progression shape of other $\pi$-conjugated systems, but notable
changes occur with decreasing temperature. At 200 K (\fg{PL}a,
bottom curve), there appears a dominant "high frequency" vibronic
progression of $\hbar\w_{H}$\smq0.17 eV (\smq1370 $cm^{-1}$),
which, however, does not match any of the Raman modes
(\fg{Raman}), and is not even in the vicinity of highly coupled
modes (Table I). The highest energy peak (marked "0") is the
fundamental optical transition ($1A_g$$\rightarrow$$1B_u$, at 3.15
eV), whereas the lower energy peaks (marked "1,2,3") are the
higher order vibronic side bands. As the temperature is lowered,
however, the high frequency progression is modulated (\fg{PL}a, 20
K) by a different "low frequency" progression of
$\hbar\w_{L}$\smq17-19 meV.  We denote this modulated vibronic
structure by $k$,$n$ ($k$=0,1,2,3, $n$=0,1,2,...), where $k$ ($n$)
is the order of the high (low) frequency modulation. The
appearance of these vibronic replicae is further illustrated in
\fg{PL}b, where we show the PL Fourier transform, namely, $|\int
{\rm PL}(E) exp(-iEt/\hbar) dE|$. At 200 K, only highly damped
short period (\smq0.02 ps) oscillations are observed. At 20 K, on
the contrary, there is considerably less damping and the short
period oscillations pattern is modulated by a long period
(\smq0.22 ps) component.

\npi The  PL spectra of DSB single crystal at 4 K and 200 K are
shown in \fg{PL}c. The overall vibronic structure is very similar
to that of the solution: a single frequency vibronic progression
at 200 K is modulated by a much lower frequency at 4 K. The main
difference between the solution and single crystal PL spectra is
the intensity of the fundamental ("0") transition: in DSB crystals
it is unobservable at 200 K, gradually increasing as the
temperature is lowered \cite{wdvf03}, reaching approximately
\smq15\% of its solution relative strength at 4 K. The suppressed
strength of the "0" transition is the result of the crystal
symmetry \cite{span01}.

 \npi When trying to account for the PL emission spectrum
of a multi-vibrational system, such as DSB, in relation with its
Raman spectrum, it is useful to employ time dependent analysis
rather than the often used sum-over-states Franck-Condon approach
\cite{hst82}. In general, the emission spectrum $F(E)$ ($E$ is the
photon energy) is given  \cite{nta80} by the Fourier transform of
the time dependent auto-correlation function of the transition
dipole moment, $f(t)$,
 \beq
F(E)=\int_{-\infty}^{\infty}f(t) e^{iEt/\hbar}dt~. \label{1}
 \eeq

\begin{figure}
\begin{center}
\includegraphics[width=6.cm]{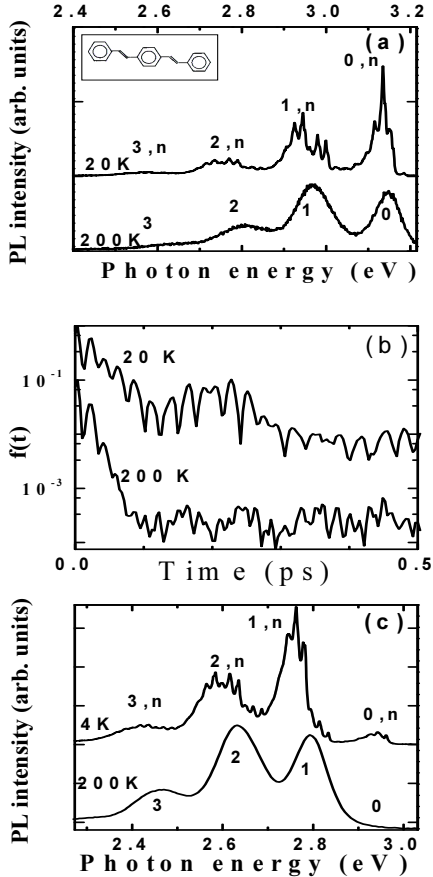}
\caption{PL spectroscopy of DSB. (a) Solution spectra at 20 K and
200 K. The indices 0 and 1,2,3 in the 200 K spectrum denote the
fundamental and vibronic replica transitions, respectively. The
pairs of indices k,n (k=0,1,2,3) in the 20 K spectrum denote the
complex modulated vibronic structure (see text). Inset: Chemical
structure of DSB. (b) Fourier transform of the PL spectra shown in
(a). Note the fast decrease at 200 K and the long period
modulation at 20 K. The 200 K curve is vertically displaced for
clarity. (c) Same as in (a) but for a DSB single crystal. Note the
missing fundamental transition at 200 K and its finite, but small,
intensity at 4 K.}
\end{center}
\end{figure}

\vspace{-0.7cm}

 Following Ref. \cite{nta80}, we write the correlation function
for a multi-mode system in the harmonic approximation and linear
electron-phonon coupling as:{{
\begin{eqnarray}
 f(t)=|P|^2e^{-iE_0t/\hbar-S+S_+(t)+S_-(t)}e^{-\Gamma |t|}~;~~~\notag \\
  S_{\pm}(t)=\sum_jS_jw_j^{\pm}e^{\mp i\w_jt}~;
 S=\sum_jS_j(2n_j+1)~,\label{3}
 \end{eqnarray}}}
\hspace{-2mm} where $P$ is the dipole matrix element for the
relevant optical transition and $E_0$ is the bare optical
transition energy.
 {{In \Eq{3}, $j$ is the mode index,
 $w_j^+=w_j^-+1=n_j+1$, where  $n_j$=1/[$exp(\hbar\w_j/k_BT)-1$]
 is the mode occupation number at temperature T,}}
$S_j$=$\w_j\D_j^2/2\hbar$, and $\D_j$ is its equilibrium (normal
coordinate) displacement in the optically excited electronic state
relative to the ground state. {{ We emphasize here that it is the
"electron temperature", $T_e$, which determines the mode
occupation, $n_j$. $T_e$ is determined by the photon excitation
energy and the electron excess energy relaxation rate and may be
considerably higher than the lattice temperature, T.}} We identify
$S$ with the system HR factor, which is the sum of the individual
HR factors modified by the temperature (\Eq{3}). {{The time
dependent term $S_{+}(t)$ in \Eq{3} is responsible for the usual
red shifted vibronic side bands in the PL spectrum, while
$S_{-}(t)$ gives rise to blue-shifted side bands (due to excited
vibrational level occupation), emphasizing the low frequency modes
at relatively high $T_e$.}} In \Eq{3}, a simple mode-independent
phenomenological damping, \G$\geq$0, is introduced; it represents
losses due to natural line broadening and/or other degrees of
freedom \cite{hst82}.
 Note
that $f(t)$ in \Eq{3} is a product of various periodical
functions, $exp(-i\w_jt)$, each with a different period and
amplitude. Thus, the effect of finite damping is to limit the
effective time domain in the integral (\Eq{1}), emphasizing
partial periodic recurrences in $f(t)$ that result in apparent
vibronic frequencies in the PL spectrum. The apparent mode (APM)
frequencies need not be equal to one of the observed Raman modes,
but are related to them in a non-trivial way \cite{ttsh83}. An
example of this effect is shown in \fg{model}a, where $f(t)$,
calculated using the 11-mode system of DSB given in Table I, is
plotted on a logarithmic scale. It is clearly seen, that even for
low damping (\fg{model}a, top) $f(t)$ is dominated by two APMs: a
high frequency (short period), APM$_H$, modulated by a low
frequency (long period), APM$_L$. The main effect of the damping
is the faster decrease of the low frequency component
(\fg{model}a, bottom).

 \npi The HR factors that determine the vibronic
structure are closely related to the measured Raman spectrum,
since the Raman process is enabled by the electron-phonon
interaction.  The T=0 intensity ($I_j^0$) of each Raman line
 measures its excited state
displacement
 \cite{hst82}: $I_j^0\propto\w_j^3\D_j^2$. We
then have \cite{nta80} $S_j$$\equiv$$\w_j\D_j^2/2\hbar\propto
I_j^0/\w_j^2$ and,
 \beq
 S_j/S(0)=[\sum_{j'}(I_{j'}^0/I_{j}^0)
 (\w_j/\w_{j'})^2]^{-1}~,\label{4}
 \eeq
where $S(0)$ is $S$ at T=0 K. We note that the $\w_j^{-2}$ factor
in $S_j$ emphasizes the lower frequency Raman modes in the PL
spectrum. This can be seen in Table I, where we list for DSB
$S_j/S(0)$, calculated using \Eq{4} and the Raman data
(\fg{Raman}): the lowest frequency mode has the largest HR factor,
although its intensity is \smq1\% of the strongest line! { We
conclude then: (1) The apparent modes are the result of a
"weighted beating" of all Raman frequencies. (2) The frequencies
of the apparent modes are not, in general, the simple sum or
difference of the Raman frequencies, and cannot therefore be
predicted {{a priori}}. (3) The relative intensities of the
apparent vibronic structure of the PL spectrum is solely
determined by the experimentally measured pre-resonant Raman
spectrum. (4) The only fitting parameters needed are the overall
damping and the absolute magnitude of the total HR factor. }

\vspace{-0.5cm}

\[
\begin{array}{c|cccccc} \hline \hline
~j&1&2&3&4&5&6\\
 ~\nu_j~(cm^{-1}) &131&261&640 &873&1000 &1181 \\
 ~I_j/I_{10}(\%)&1.16&0.45&0.71 &0.77&6.7 &50 \\
~S_j/S&0.35&0.039&0.01 &0.01&0.035 &0.19 \\ \hline
 ~j&7&8&9&10&11&~\\
~\nu_j~(cm^{-1}) &1330&1452&1561 &1591&1635& \\
 ~I_j/I_{10}(\%)&19&2.7&13 &100&42& \\
~S_j/S&0.058&0.007&0.027 &0.20&0.078& \\
\hline \hline
\end{array} \]

Table I. {\small The most intense Raman lines of crystalline DSB
at 10 K. $\nu_j$, $I_j/I_{10}$ and $S_j/S$, respectively, denote
the measured peak frequency, relative integrated intensity and the
calculated relative HR factor (\Eq{4}, for each mode.}

\begin{figure}
\begin{center}
\includegraphics[width=6.cm,angle=-90]{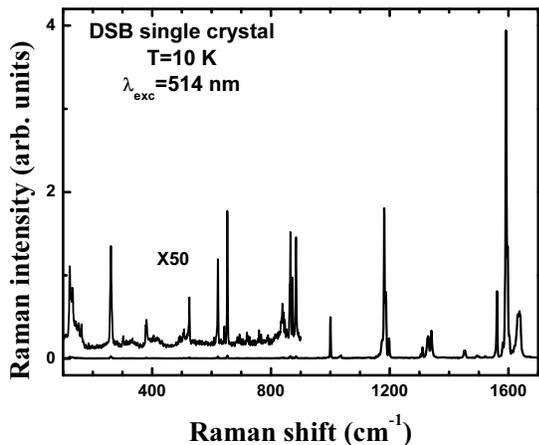}
\caption{Raman spectrum of DSB single crystal at pre-resonant
conditions.}
\end{center}
\end{figure}

\vspace{-0.5cm}

\npi The first excited state of isolated DSB molecules is
optically allowed, making them strongly luminescent and useable as
active media in light emitting devices \cite{Hadz}. Typically, the
PL spectrum of isolated DSB molecules consists of the fundamental
(0-0) optical transition and a single frequency phonon side bands
replica series. However, in DSB and other
p-oligophenylene-vinylenes, as well as oligothiophens
\cite{mcsb02}, in the form of solid films and crystals, the 0-0
band is strongly suppressed, whereas the phonon replicae retain
their intensity and dominate the PL spectrum \cite{oegt96,lmmh00}.
{{The 0-0 PL band in films and crystals is forbidden due to
Davidov splitting associated with H-aggregates.}} The 0-0
intensity depends on the film morphology (or aggregate size)
and/or crystal purity. In an inhomogeneous solid sample, the
resulting emission spectrum may be composed of contributions from
several domains, with various 0-0 to 0-n intensity ratio. We
therefore discuss separately the emission spectra of solution and
crystalline DSB.

\vspace{-7mm}

\begin{figure}
\begin{center}
\includegraphics[width=14.cm,angle=-90]{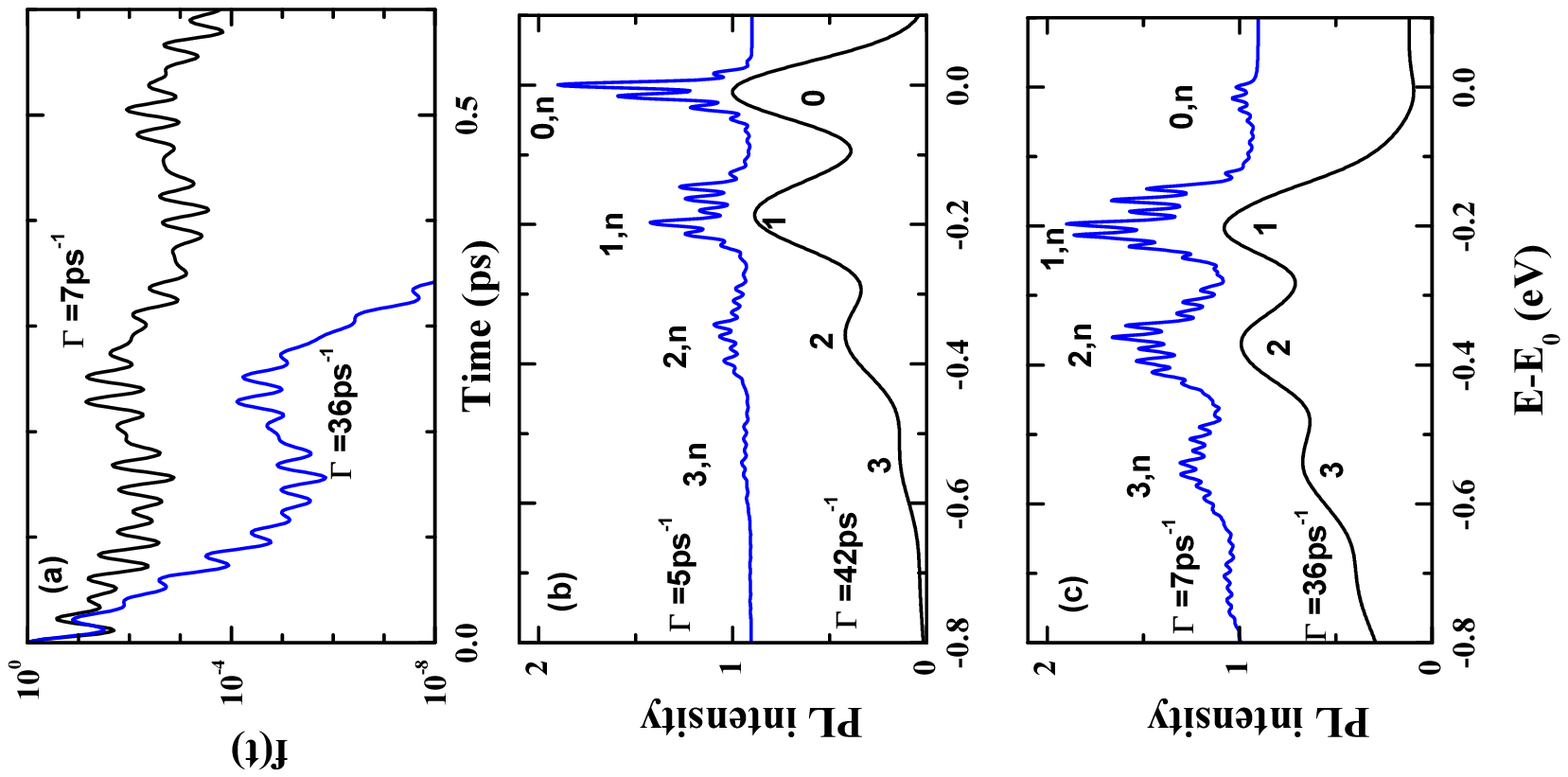}
\caption{Model calculations of DSB PL spectra using the 11-mode
system of Table I for \G\ as shown. (a) Normalized dipole
auto-correlation function $f(t)$ (\Eq{3}) for S(0)=2.5, $T_e$=4
(200) K for the upper (lower) curve; note the log scale. (b) PL
spectra obtained by the Fourier transform of $f(t)$ similar to
that in (a), but for S(0)=1.7, $T_e$=150 (200) K for the upper
(lower) curve. $E_0$ is the energy of the fundamental transition.
(c) PL spectra obtained by the Fourier transform of $f(t)$ for
parameters as in (a), but for the totally suppressed (bottom) and
85\% suppressed (top) fundamental transition.}
\end{center}
\end{figure}

\vspace{-7mm}

\npi {{Using the data of Table I we show in \fg{model}a the dipole
auto-correlation function, $f(t)$, generated for low and high
damping. It is visually striking that the 11-mode system is
dominated by only two APM: a short period mode modulated by a long
period mode. Moreover, the frequency associated with these two
APMs does not coincide with any DSB normal mode. The data given in
Table I is also used to generate the PL spectra shown in
\fg{model}b. Here, the values of the electron temperature, $T_e$,
HR factor $S$ and damping $\G$ were chosen to best fit the frozen
solution experimental data at low and high lattice temperatures, T
(\fg{PL}a). The higher damping spectrum that presumably occurs at
T=200 K ($\G=42 ps^{-1}$, \fg{model}b) shows a vibronic
progression dominated by a single frequency, APM$_H$=173 meV
(\smq1400 cm$^{-1}$). These progression peaks are marked as
"1,2,3". This is in excellent agreement with the data at 200 K
(\fg{PL}a, bottom curve). For this case we chose $T_e=T$, since
the relevant modes are at high frequencies. At lower lattice
temperatures we expect the damping to decrease. Consequently, the
low frequency APM is less suppressed, making the PL spectrum
sensitive to the actual value of $T_e$. At low \G\ values, there
appears a low frequency modulation (with APM$_L$\smq17 meV) of the
high frequency vibronic series, as seen in \fg{model}b (top
curve). The combined progression peaks are denoted as (k,n), where
k=0,1,2,3 denotes the APM$_H$ progression and n=0,1,2... denotes
the APM$_L$ modulation of APM$_H$. The PL spectrum in \fg{model}b
(top curve) was calculated using $T_e=150$K; it shows two blue
shifted peaks, in very good agreement with the experimental data
at T=20 K (\fg{PL}a, top curve). We thus conclude that due to the
non-resonant PL excitation, $T_e>T$.}}

\npi The crystalline PL spectra show similar characteristics to
the solution spectra; i.e., at high temperatures is dominated by a
single APM, whereas at low temperatures this structure is
modulated by a low frequency APM. As for the solution spectra,
this behavior naturally results from a decreased damping at low
temperatures. However, the crystalline spectra reveal an
additional feature not observed in solutions.

\npi It is seen in the PL spectra of crystalline DSB (\fg{PL}c),
that the band centered at \smq2.95 eV appears at T=4 K but not at
T=200 K. Measurements at intermediate temperatures \cite{wdvf03}
reveal that its intensity monotonically decreases with increasing
temperature; above T=150 K it cannot be observed any longer. We
interpret this band as the fundamental optical transition that is
suppressed by the crystal symmetry. However, the vibronic
structure below \smq2.95 eV is not affected \cite{span01}. In
order to quantitatively account for the temperature dependent
crystalline PL spectra, we have allowed the "0" transition to vary
independently of all other phonon mediated transitions.

\npi First, we recognize that the $S_+(t)$ terms in \Eq{3} are
responsible for all the red shifted spectral features associated
with the electron-phonon interaction. The time independent term
$S$ in \Eq{3} gives the intensity of the fundamental optical
transition in the presence of the coupled vibrational modes.
Second, assuming that the fundamental optical transition has a
small finite width of $\hbar\G_0$, we calculate its intensity
(denoted $I_0(E)$) using \Eq{3} while taking into account only
those modes whose frequency is smaller than $\G_0^{-1}$. We then
let $I_0$ to be partially suppressed, and write for the
crystalline PL spectrum,
 \beq
F_{cryst}(E)=F(E)-\alpha I_0(E)~,\label{5}
 \eeq
where $F(E)$ is given by \Eq{1}, and $\alpha$ is the suppression
parameter controlling $I_0$ in the crystalline spectrum. Using
\Eq{5},  the data of Table I, and $\hbar\G_0$=4 meV we show in
\fg{model}c a high temperature (200K, \G=36 ps$^{-1}$) PL spectrum
with totally suppressed fundamental transition, namely $\alpha$=1.
This is in excellent agreement with the T=200 K crystalline
spectrum (\fg{PL}c).

\npi Allowing now for a non-zero $I_0$ intensity ($\alpha$$<$1)
together with a smaller \G, we can account for the  finite
intensity fundamental transition and the low frequency modulation
of the high frequency APM observed in crystalline PL spectra at
low temperatures. An illustrative example is shown in \fg{model}c
for the T=4 K spectrum, where $I_0$ is 85\% suppressed and
APM$_L$\smq17 meV; again, in very good agreement with the 4 K
crystalline spectrum (\fg{PL}c).

\npi In summary, we showed that the apparent low frequency
vibronic side band in the crystalline PL spectrum is inherent to
the DSB chromophore; it is not related to the strength of $I_0$,
as previously suggested for oligothiophene crystals \cite{mcsb02}.
{{We interpret the blue shifted  small peaks above the 0-0
transition, in the frozen solution spectrum, as due to a "hot
luminescence" process. }}Using a time dependent model with
temperature dependent damping, we showed that the many-mode DSB
system is effectively transformed into two- then to a single
"apparent" mode as damping increases. In general, increased
damping in multi-vibrational mode $\pi$-conjugated systems results
in effective elimination of vibrational modes from the emission
and absorption spectra and the eventual appearance of a nearly
regularly spaced progression at an apparent frequency. Knowing the
Raman spectrum, it is possible then to account in detail for the
emission and absorption spectra of $\pi$-conjugated systems.

{\bf Acknowledgments}--Supported by DOE FG-02-04 ER46109 and
Israel Science Foundation 735/04.

\end{multicols}
\end{document}